\documentclass[10pt,a4paper,onecolumn]{article}
\usepackage{marginnote}
\usepackage{graphicx}
\usepackage[rgb,dvipsnames]{xcolor}
\usepackage{authblk,etoolbox}
\usepackage{titlesec}
\usepackage{calc}
\usepackage{tikz}
\usepackage[pdfa]{hyperref}
\usepackage{hyperxmp}

\usepackage[backend=biber, style=numeric]{biblatex}

\hypersetup{%
    unicode=true,
    pdfapart=3,
    pdfaconformance=B,
    pdftitle={TelescopeML -- I. An End-to-End Python Package for Interpreting Telescope Datasets through Training Machine Learning Models, Generating Statistical Reports, and Visualizing Results},
    pdfauthor={Ehsan (Sam) Gharib-Nezhad, Natasha E. Batalha, Hamed Valizadegan, Miguel J. S. Martinho, Mahdi Habibi, Gopal Nookula},
    pdfpublication={Journal of Open Source Software},
    pdfpublisher={Open Journals},
    pdfissn={2475-9066},
    pdfpubtype={journal},
    pdfvolumenum={},
    pdfissuenum={},
    pdfdoi={10.21105/joss.06346},
    pdfcopyright={Copyright (c) 2024, Ehsan (Sam) Gharib-Nezhad, Natasha E. Batalha, Hamed Valizadegan, Miguel J. S. Martinho, Mahdi Habibi, Gopal Nookula},
    pdflicenseurl={http://creativecommons.org/licenses/by/4.0/},
    colorlinks=true,
    linkcolor=[rgb]{0.0, 0.5, 1.0},
    citecolor=Blue,
    urlcolor=[rgb]{0.0, 0.5, 1.0},
    breaklinks=true
}

\immediate\pdfobj stream attr{/N 3} file{sRGB.icc}
\pdfcatalog{%
  /OutputIntents [
    <<
      /Type /OutputIntent
      /S /GTS_PDFA1
      /DestOutputProfile \the\pdflastobj\space 0 R
      /OutputConditionIdentifier (sRGB)
      /Info (sRGB)
    >>
  ]
}

\usepackage{caption}
\usepackage{orcidlink}
\usepackage{tcolorbox}
\usepackage{amssymb,amsmath}
\usepackage{ifxetex,ifluatex}
\usepackage{seqsplit}
\usepackage{xstring}

\usepackage{float}
\let\origfigure\figure
\let\endorigfigure\endfigure
\renewenvironment{figure}[1][2] {
    \expandafter\origfigure\expandafter[H]
} {
    \endorigfigure
}

\usepackage{fixltx2e} 

\makeatletter
 \let\@cite@ofmt\@firstofone
 \def\@biblabel#1{}
 \def\@cite#1#2{{#1\if@tempswa , #2\fi}}
\makeatother
\newlength{\cslhangindent}
\newlength{\csllabelwidth}
 {\begin{list}{}{%
  \setlength{\itemindent}{0pt}
  \setlength{\leftmargin}{0pt}
  \setlength{\parsep}{0pt}
  \ifodd #1
   \setlength{\leftmargin}{\cslhangindent}
   \setlength{\itemindent}{-1\cslhangindent}
  \fi
  \setlength{\itemsep}{#2\baselineskip}}}
 {\end{list}}
\usepackage{calc}

\usepackage[top=3.5cm, bottom=3cm, right=1.5cm, left=1.0cm,
            headheight=2.2cm, reversemp, includemp, marginparwidth=4.5cm]{geometry}

\titleformat{\section}
  {\normalfont\sffamily\Large\bfseries}
  {}{0pt}{}
\titleformat{\subsection}
  {\normalfont\sffamily\large\bfseries}
  {}{0pt}{}
\titleformat{\subsubsection}
  {\normalfont\sffamily\bfseries}
  {}{0pt}{}
\titleformat*{\paragraph}
  {\sffamily\normalsize}

\usepackage{fancyhdr}
\makeatletter
\let\ps@plain\ps@fancy
\fancyheadoffset[L]{4.5cm}
\fancyfootoffset[L]{4.5cm}

\definecolor{linky}{rgb}{0.0, 0.5, 1.0}

\newtcolorbox{repobox}
   {colback=red, colframe=red!75!black,
     boxrule=0.5pt, arc=2pt, left=6pt, right=6pt, top=3pt, bottom=3pt}

\newcommand{\ExternalLink}{%
   \tikz[x=1.2ex, y=1.2ex, baseline=-0.05ex]{%
       \begin{scope}[x=1ex, y=1ex]
           \clip (-0.1,-0.1)
               --++ (-0, 1.2)
               --++ (0.6, 0)
               --++ (0, -0.6)
               --++ (0.6, 0)
               --++ (0, -1);
           \path[draw,
               line width = 0.5,
               rounded corners=0.5]
               (0,0) rectangle (1,1);
       \end{scope}
       \path[draw, line width = 0.5] (0.5, 0.5)
           -- (1, 1);
       \path[draw, line width = 0.5] (0.6, 1)
           -- (1, 1) -- (1, 0.6);
       }
   }

\patchcmd{\@maketitle}{center}{flushleft}{}{}
\patchcmd{\@maketitle}{center}{flushleft}{}{}
\patchcmd{\@maketitle}{\LARGE}{\LARGE\sffamily}{}{}
\def\maketitle{{%
  
  \AB@maketitle}}
\renewcommand\AB@affilsepx{ \protect\Affilfont}
\renewcommand\AB@affilnote[1]{{\bfseries #1}\hspace{3pt}}
\renewcommand{\affil}[2][]%
   {\newaffiltrue\let\AB@blk@and\AB@pand
      \if\relax#1\relax\def\AB@note{\AB@thenote}\else\def\AB@note{#1}%
        \setcounter{Maxaffil}{0}\fi
        \begingroup
        \let\href=\href@Orig
        \let\protect\@unexpandable@protect
        \def\thanks{\protect\thanks}\def\footnote{\protect\footnote}%
        \@temptokena=\expandafter{\AB@authors}%
        {\def\\{\protect\\\protect\Affilfont}\xdef\AB@temp{#2}}%
         \xdef\AB@authors{\the\@temptokena\AB@las\AB@au@str
         \protect\\[\affilsep]\protect\Affilfont\AB@temp}%
         \gdef\AB@las{}\gdef\AB@au@str{}%
        {\def\\{, \ignorespaces}\xdef\AB@temp{#2}}%
        \@temptokena=\expandafter{\AB@affillist}%
        \xdef\AB@affillist{\the\@temptokena \AB@affilsep
          \AB@affilnote{\AB@note}\protect\Affilfont\AB@temp}%
      \endgroup
       \let\AB@affilsep\AB@affilsepx
}
\makeatother

\renewcommand\Affilfont{\sffamily\small\mdseries}
\ifnum 0\ifxetex 1\fi\ifluatex 1\fi=0 
  \usepackage[T1]{fontenc}
  \usepackage[utf8]{inputenc}

\else 
  \ifxetex
    \usepackage{mathspec}
    \usepackage{fontspec}

  \else
    \usepackage{fontspec}
  \fi
  \defaultfontfeatures{Scale=MatchLowercase}
  \defaultfontfeatures[\sffamily]{Ligatures=TeX}
  \defaultfontfeatures[\rmfamily]{Ligatures=TeX,Scale=1}

\fi
\IfFileExists{upquote.sty}{\usepackage{upquote}}{}
\IfFileExists{microtype.sty}{%
\usepackage{microtype}
\UseMicrotypeSet[protrusion]{basicmath} 
}{}

\PassOptionsToPackage{usenames,dvipsnames}{color} 
\ifLuaTeX
\usepackage[bidi=basic]{babel}
\else
\usepackage[bidi=default]{babel}
\fi
\babelprovide[main,import]{american}

\def\languageshorthands#1{}

\usepackage{graphicx,grffile}
\makeatletter
\def\maxwidth{\ifdim\Gin@nat@width>\linewidth\linewidth\else\Gin@nat@width\fi}
\def\maxheight{\ifdim\Gin@nat@height>\textheight\textheight\else\Gin@nat@height\fi}
\makeatother
\setkeys{Gin}{width=\maxwidth,height=\maxheight,keepaspectratio}
\IfFileExists{parskip.sty}{%
\usepackage{parskip}
}{
\setlength{\parindent}{0pt}
\setlength{\parskip}{6pt plus 2pt minus 1pt}
}
\ifx\paragraph\undefined\else
\let\oldparagraph\paragraph
\renewcommand{\paragraph}[1]{\oldparagraph{#1}\mbox{}}
\fi
\ifx\subparagraph\undefined\else
\let\oldsubparagraph\subparagraph
\renewcommand{\subparagraph}[1]{\oldsubparagraph{#1}\mbox{}}
\fi
\ifLuaTeX
  \usepackage{selnolig}  
\fi

\title{\texttt{TelescopeML} -- I. An End-to-End Python Package for Interpreting Telescope Datasets through Training Machine Learning Models, Generating Statistical Reports, and Visualizing Results}

\author[1, 2]{Ehsan (Sam) Gharib-Nezhad\thanks{ORCID: 0000-0002-4088-7262}}
\author[1]{Natasha E. Batalha\thanks{ORCID: 0000-0003-1240-6844}}
\author[3, 4]{Hamed Valizadegan\thanks{ORCID: 0000-0001-6732-0840}}
\author[3, 4]{Miguel J. S. Martinho\thanks{ORCID: 0000-0002-2188-0807}}
\author[5]{Mahdi Habibi\thanks{ORCID: 0000-0001-8530-7746}}
\author[6]{Gopal Nookula}

\affil[1]{Space Science and Astrobiology Division, NASA Ames Research Center, Moffett Field, CA, 94035 USA}
\affil[2]{Bay Area Environmental Research Institute, NASA Research Park, Moffett Field, CA 94035, USA}
\affil[3]{Universities Space Research Association (USRA), Mountain View, CA 94043, USA}
\affil[4]{Intelligent Systems Division, NASA Ames Research Center, Moffett Field, CA 94035, USA}
\affil[5]{Institute for Radiation Physics, Helmholtz-Zentrum Dresden-Rossendorf, Dresden 01328, Germany}
\affil[6]{Department of Computer Science, University of California, Riverside, Riverside, CA 92507 USA}

\begin{document}
\maketitle

\marginpar{

  \begin{flushleft}
  \sffamily\small

  {\bfseries DOI:} \href{https://doi.org/10.21105/joss.06346}{\color{linky}{10.21105/joss.06346}}

  \vspace{2mm}
    {\bfseries Software}
  \begin{itemize}
    \setlength\itemsep{0em}
    \item \href{https://github.com/openjournals/joss-reviews/issues/6346}{\color{linky}{Review}} \ExternalLink
    \item \href{https://github.com/EhsanGharibNezhad/TelescopeML}{\color{linky}{Repository}} \ExternalLink
    \item \href{https://zenodo.org/records/11553655}{\color{linky}{Archive}} \ExternalLink
  \end{itemize}

  \vspace{2mm}
  
    \par\noindent\hrulefill\par

  \vspace{2mm}

  {\bfseries Editor:} \href{https://github.com/dfm}{Dan
Foreman-Mackey} \ExternalLink
  \,\orcidlink{0000-0002-9328-5652} \\
  \vspace{1mm}
    {\bfseries Reviewers:}
  \begin{itemize}
  \setlength\itemsep{0em}
    \item \href{https://github.com/oparisot}{@oparisot}
    \item \href{https://github.com/mwalmsley}{@mwalmsley}
    \end{itemize}
    \vspace{2mm}
  
    {\bfseries Submitted:} 29 November 2023\\
    {\bfseries Published:} 18 July 2024

  \vspace{2mm}
  {\bfseries License}\\
  Authors of papers retain copyright and release the work under a Creative Commons Attribution 4.0 International License (\href{https://creativecommons.org/licenses/by/4.0/}{\color{linky}{CC BY 4.0}}).

  \end{flushleft}
}

\section*{Summary}

We are on the verge of a revolutionary era in space exploration, thanks to advancements in telescopes such as the James 
Webb Space Telescope (\textit{JWST}). High-resolution, high signal-to-noise spectra from exoplanet and brown dwarf atmospheres have been 
collected over the past few decades, requiring the development of accurate and reliable pipelines and tools for their analysis. 
Accurately and swiftly determining the spectroscopic parameters from the observational spectra of these objects is 
crucial for understanding their atmospheric composition and guiding future follow-up observations. \texttt{TelescopeML} is a 
Python package developed to perform three main tasks: 
\begin{enumerate}
    \item Process the synthetic astronomical datasets for training a CNN model and prepare the observational dataset for later use for prediction;
    \item Train a CNN model by implementing the optimal hyperparameters; and 
    \item Deploy the trained CNN models on the actual observational data to derive the output spectroscopic parameters.
\end{enumerate}

The implications and scientific outcomes from the trained CNN models and this package are under revision for The 
Astrophysical Journal under the title \textit{TelescopeML – II: Convolutional Neural Networks for Predicting Brown Dwarf 
Atmospheric Parameters}.

\section*{Statement of Need}

We are in a new era of space exploration, thanks to advancements in ground- and space-based 
telescopes, such as the James Webb Space Telescope \cite{JWST2023PASP} and CRIRES. These remarkable instruments collect high-resolution, high-signal-to-noise spectra 
from extrasolar planets \cite{Alderson2023Nature}, and brown dwarfs \cite{Miles2023ApJ} atmospheres. Without accurate interpretation of this data, the main objectives 
of space missions will not be fully accomplished. Different analytical and statistical methods, such as the chi-squared-test, 
Bayesian statistics, and radiative-transfer atmospheric modeling packages have been developed 
\cite{batalha2019picaso, MacDonald2023} to interpret the spectra. They utilize either forward- and/or retrieval-radiative transfer modeling to analyze the spectra and 
extract physical information, such as atmospheric temperature, metallicity, carbon-to-oxygen ratio, and surface gravity 
\cite{line2014systematic, Iyer2023Sphinx, Marley2015}. These atmospheric models rely on generating the physics and chemistry of these atmospheres for a wide range of thermal structures 
and compositions. In addition to Bayesian-based techniques, machine learning and deep learning methods have been developed in recent years 
for various astronomical problems, including confirming the classification of light curves for 
exoplanet validation \cite{Valizadegan2022}, recognizing molecular features \cite{Zingales2018ExoGAN} as well as interpreting brown dwarfs spectra using Random Forest technique 
\cite{Lueber2023RandomForesr_BDs}. Here, we present one of the first applications of deep learning and convolutional neural networks on the interpretation of brown dwarf 
atmospheric datasets. The configuration of a CNN and the key concepts can be found in \cite{Goodfellow_2016DeepLearning, KIRANYAZ2021}.

With the continuous observation of these objects and the increasing amount of data, there is a 
critical need for a systematic pipeline to quickly explore the datasets and extract important physical parameters from them. In the future, we can expand our pipeline to exoplanet atmospheres, and use it to provide insights about the diversity of exoplanets and brown dwarfs' 
atmospheric compositions. Ultimately, \texttt{TelescopeML} will help facilitate the long-term analysis of this data in research. \texttt{TelescopeML}
is an ML Python package with Sphinx-ed user-friendly documentation that provides both trained ML models and ML tools 
for interpreting observational data captured by telescopes.

\section*{Functionality and Key Features}

\texttt{TelescopeML} is a Python package comprising a series of modules, each equipped with specialized machine learning and 
statistical capabilities for conducting Convolutional Neural Networks (CNN) or Machine Learning (ML) training on datasets 
captured from the atmospheres of extrasolar planets and brown dwarfs. The tasks executed by the \texttt{TelescopeML} modules are 
outlined below and visualized in the following figure:

\begin{itemize}
    \item \textbf{DataMaster module}: Performs various tasks to process the datasets, including:
    \begin{itemize}
        \item Load the training dataset (i.e., atmospheric fluxes) in CSV format
        \item Split the dataset into training, validation, and test sets to pass it to the CNN model
        \item Scale/normalize the dataset column-wise or row-wise
        \item Visualize the training sets in each of the processing steps for more insights 
        \item Perform feature engineering by extracting the Min and Max values from each flux to improve the ML training performance
    \end{itemize}
  
    \item \textbf{DeepTrainer module}: Utilizes different methods/packages such as TensorFlow to:
    \begin{itemize}
        \item Load the processed dataset from the \textbf{DataMaster} module
        \item Build Convolutional Neural Networks (CNNs) model using the tuned hyperparameters
        \item Fit/train the CNN models given the epochs, learning rate, and other parameters
        \item Visualize the loss and training history, as well as the trained model's performance
    \end{itemize}
  
    \item \textbf{Predictor module}: Implements the following tasks to predict atmospheric parameters: 
    \begin{itemize}
        \item Perform Scale/normalize processes on the observational fluxes
        \item Deploy the trained CNNs model 
        \item Predict atmospheric parameters, i.e., effective temperature, gravity, carbon-to-oxygen ratio, and metallicity 
        \item Visualize the processed observational dataset and the uncertainty in the predicted results
    \end{itemize}
  
    \item \textbf{StatVisAnalyzer module}: Provides a set of functions to perform the following tasks: 
    \begin{itemize}
        \item Explore and process the synthetic datasets
        \item Perform the chi-square test to evaluate the similarity between two datasets
        \item Calculate confidence intervals and standard errors
    \end{itemize}
\end{itemize}

\begin{figure}[h!]
    \centering
    \includegraphics[height=900pt]{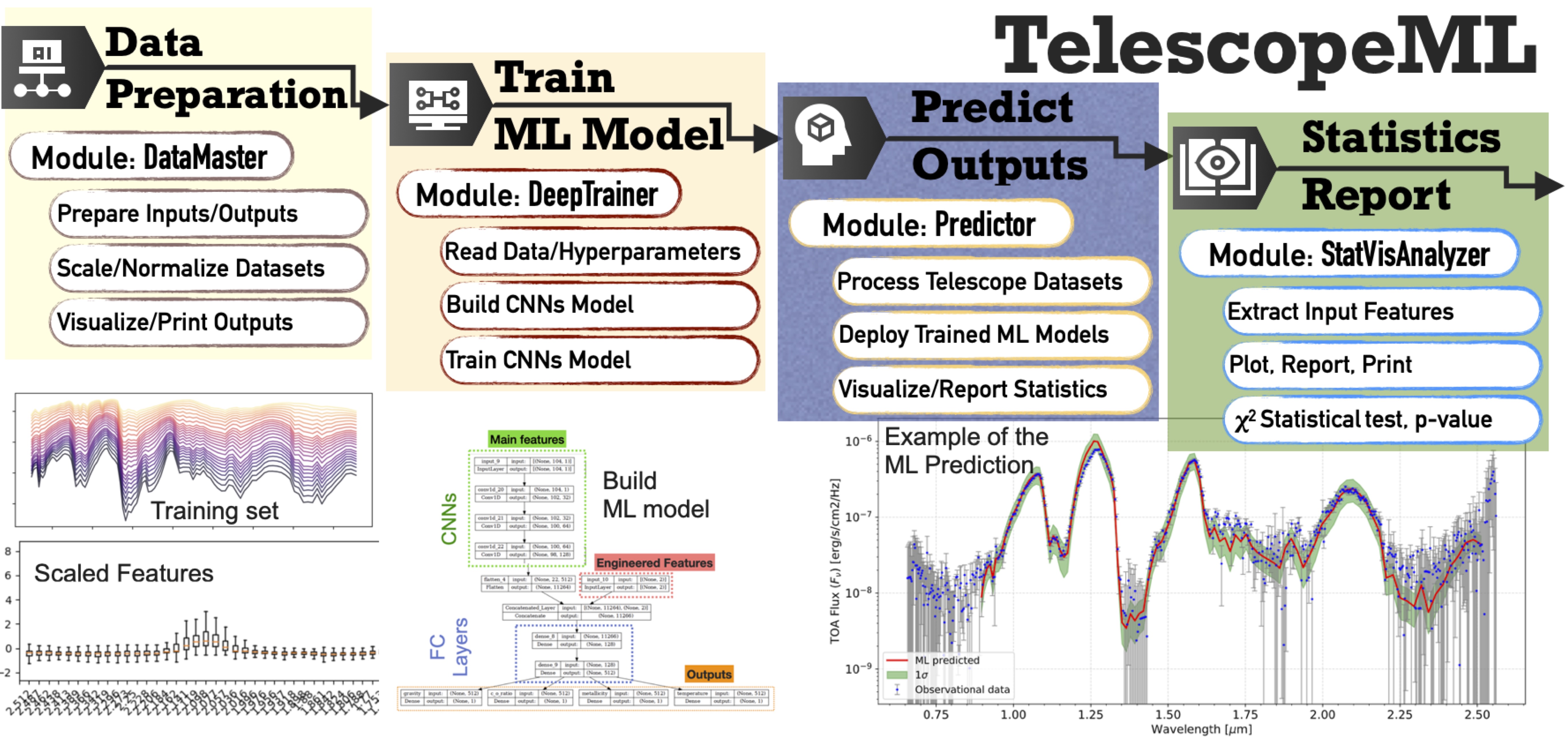}
    \caption{TelescopeML main modules to manipulate the training example, build the ML model, train and tune it, and ultimately extract the target features from the observational data.}
    \label{fig:telescope_ml_modules}
\end{figure}

\section*{Details on the synthetic dataset}

The training dataset (or synthetic spectra) in this study is computed using the open-source atmospheric radiative 
transfer Python package, \href{https://natashabatalha.github.io/picaso/}{\texttt{PICASO}} \cite{batalha2019picaso}, based on the 
\texttt{Sonora-Bobcat} model grid generated for cloudless brown dwarf atmospheres by \cite{marley2021sonora}. This set encompasses  30,888 synthetic spectra, each including 104 wavelengths (i.e., 0.897, 0.906, ..., 2.512 $\mu$m) and their corresponding flux 
values. Each of these spectra has four output variables attached to it: effective temperature, gravity, carbon-to-oxygen ratio, 
and metallicity. These synthetic spectra are utilized to interpret observational datasets and derive these four atmospheric parameters.
An example of the synthetic and observational dataset is shown in the following figure.

\section*{Details on the CNN Methodology for Multi-output Regression Problem}

Each row in the synthetic spectra has 104 input variables. The order of these data points and their magnitude are crucial 
to interpret the telescope data. For this purpose, we implemented a Convolutional Neural Network (CNN) method with 1-D convolutional 
layers. CNN is a powerful technique for this study because it extracts the dominant features from these spectra and then passes them 
to the fully connected hidden layers to learn the patterns. The output layer predicts the four atmospheric targets.
An example of the CNN architecture is depicted in the following figure:

\section*{Documentation}

\texttt{TelescopeML} is available and being maintained as a GitHub repository at
\url{https://github.com/EhsanGharibNezhad/TelescopeML}. Online 
documentation is hosted with \textit{Sphinx} using \textit{ReadtheDocs} tools and includes several instructions and tutorials 
as follows:

\begin{itemize}
    \item \textbf{Main page}: \url{https://ehsangharibnezhad.github.io/TelescopeML/}
    \item \textbf{Installation}: \url{https://ehsangharibnezhad.github.io/TelescopeML/installation.html}
    \item \textbf{Tutorials and examples}: \url{https://ehsangharibnezhad.github.io/TelescopeML/tutorials.html}
    \item \textbf{The code}: \url{https://ehsangharibnezhad.github.io/TelescopeML/code.html}
\end{itemize}

\section*{Users and Future Developments}

Astrophysicists with no prior machine learning knowledge can deploy the \texttt{TelescopeML} package and download the 
pre-trained ML or CNN models to interpret their observational data. In this scenario, pre-trained ML models, 
as well as the PyPI package, can be installed and deployed following the online instructions. Tutorials in the 
Sphinx documentation include examples for testing the code and also serve as a starting point. For this purpose, 
a basic knowledge of Python programming is required to install the code, run the tutorials, deploy the modules, 
and extract astronomical features from their datasets. The necessary machine learning background and a detailed 
guide for package installation, along with links to further Python details, are provided to help understand 
the steps and outputs.

Astrophysicists with machine learning expertise and data scientists can also benefit from this package by 
developing and fine-tuning the modules and pre-trained models to accommodate more complex datasets from 
various telescopes. This effort could also involve the utilization of new ML and deep learning algorithms, 
adding new capabilities such as feature engineering methods, and further optimization of hyperparameters 
using different and more efficient statistical techniques. The ultimate outcome from these two groups would 
be the creation of more advanced models with higher performance and robustness, as well as the extension of 
the package to apply to a wider range of telescope datasets.

\section*{Similar Tools}

The following open-source tools are available to either perform forward modeling ($\chi^2$-based test) or retrievals 
(based on Bayesian statistics and posterior distribution):
\begin{itemize}
    \item \href{https://starfish.readthedocs.io/en/latest/index.html}{\texttt{Starfish}} \cite{Czekala2015starfish}
    \item \href{https://gitlab.com/mauricemolli/petitRADTRANS}{\texttt{petitRADTRANS}} \cite{Molliere2019}
    \item \href{https://github.com/MartianColonist/POSEIDON}{\texttt{POSEIDON}} \cite{MacDonald2023}
    \item \href{https://github.com/ideasrule/platon}{\texttt{PLATON}} \cite{Zhang2019}
    \item \href{https://github.com/mrline/CHIMERA}{\texttt{CHIMERA}} \cite{Line2013}
    \item \href{https://github.com/ucl-exoplanets/TauREx3_public}{\texttt{TauRex}} \cite{Waldmann2015}
    \item \href{https://github.com/nemesiscode/radtrancode}{\texttt{NEMESIS}} \cite{Irwin2008}
    \item \href{https://github.com/pcubillos/pyratbay}{\texttt{Pyrat Bay}} \cite{Cubillos2021}
\end{itemize}

In addition, the following package implements random forest to predict the atmospheric parameters:
\begin{itemize}
    \item \href{https://github.com/exoclime/HELA}{\texttt{HELA}} \cite{MarquezNeila2018HELA}
\end{itemize}

\section*{Utilized Underlying Packages}

For processing datasets and training ML models in \texttt{TelescopeML}, the following software/packages are employed:
\begin{itemize}
    \item Scikit-learn \cite{scikit-learn}
    \item TensorFlow \cite{tensorflow2015-whitepaper}
    \item AstroPy \cite{astropy:2022}
    \item SpectRes \cite{SpectRes}
    \item Pandas \cite{reback2020pandas}
    \item NumPy \cite{harris2020array}
    \item SciPy \cite{2020SciPy-NMeth}
    \item Matplotlib \cite{Hunter:2007}
    \item Seaborn \cite{Waskom2021}
    \item Bokeh \cite{bokeh}
\end{itemize}
Additionally, for generating training astronomical datasets, 
\href{https://natashabatalha.github.io/picaso/}{\texttt{PICASO}} \cite{batalha2019picaso} is implemented.

\section*{Acknowledgements}

EGN and GN would like to thank OSTEM internships and funding through NASA with contract number 80NSSC22DA010.
EGN acknowledges ChatGPT 3.5 for proofreading some of the functions. EGN is grateful to Olivier Parisot and 
Mike Walmsley for helpful referee reports, and to the JOSS editorial staff, Paul La Plante and Dan Foreman-Mackey, for
their tireless efforts to encourage new people to join the open source community in astronomy.

\printbibliography

\end{document}